# Concurrent Kink and Sausage Waves in A Crescent Shaped Structure Over A Limb Prominence

Maryam Ghiasi,[1] Neda Dadashi,[2]⋆ Hossein Ebadi[1]†
[1]*Faculty of physics, University of Tabriz, Tabriz, Iran*
[2]*Department of physics, University of Zanjan, Zanjan, Iran, P.O.Box 45371-38791,*

**ABSTRACT**
A Crescent shaped prominence Structure (CS), over the solar west limb is studied using EIS/HINODE and AIA/SDO. First, the time varying positions of the top and below borders of the CS, along with its central axis are derived. Time evolutions of the Doppler shifts, and Line width of Fe XII 195.119 line are studied over the CS borders. Transverse kink oscillations are observed both in the Solar-Y direction and in the Doppler shifts over the observers' LOS. One explanation could be the oscillatory direction of the main kink wave build an angle with the observers LOS. This angle is calculated to be equal to 27 degrees for the CS top border. The main kink amplitude velocity and periods are obtained to be 5.3 km/s, and 33.4 minutes, respectively. The anti-correlation observed between the brightness and thickness of the CS (with -178.1°) suggests the presence of sausage modes with periods of 20.8 minutes. Based on the AIA imaging, it is suggested that the occurred jets and their afterward dimming are responsible to trigger the sausage mode. The average electron densities of the CS over the time of the study is obtained to be $\log(n_e)=9.3[\text{cm}^{-3}]$. The Alfvén velocity, magnetic field, and energy flux of the observed fast kink mode over the CS are estimated to be 16.7 km/s, 2.79 G, 41.93 W/m². Considering the magnetic flux conservation inside the CS, expanding the CS cross section, causes the magnetic field to decay with the rate of $4.95 \times 10^{-4}$ G/sec.

## 1 INTRODUCTION

MHD waves are believed to be one of the main candidates to transfer the energy and momentum from the lower to the upper atmospheric heights. In a cylindrical magnetic flux tube, which is a very common structure among the solar atmospheric features, three different wave modes could propagate: kink, sausage, and torsional Alfvén waves. Torsional Alfvén waves are incompressible in nature, and propagates along the magnetic field lines. The restoring force of this transverse mode is magnetic tension. Zaqarashvili (2003) stated that the torsional Alfvén waves can be detected by periodic broadening of spectral lines if the line-of-sight does not be parallel to the wave axis. Observational signatures of this mode were reported later than the fast and slow waves, because this mode does not show intensity variation and this makes their observation difficult. Jess et al. (2009) using the Swedish Solar Telescope observed this indirect signature of torsional Alfvén waves in the H$\alpha$ data in lower levels of solar atmosphere with periods of 2 to 12 minutes over a large bright point group near the disk center with area of 430,000 km². Van Doorsselaere et al. (2008) predicted these line width periodic broadenings (due to torsional Alfvén waves) to be 180 degrees anti-phase at opposite boundaries of the waveguide(Mathioudakis et al. 2012). Using high resolution imaging instruments, along with the high resolution (and sensitive) spectrometric and/or polarimetric instruments people detect signatures of the torsional Alfvén waves (Banerjee et al. 2009; Antolin & Shibata 2010; McIntosh et al. 2011; De Moortel & Nakariakov 2012; Mathioudakis et al. 2012; Jess et al. 2015; Morton et al. 2015).

Sausage modes (with an azimuthal wave-number, m = 0), are compressible waves periodically changing the density (and line intensity) of the flux tube by contraction and expansion of its cross section (Nakariakov & Verwichte 2005). This mode is detected for the first time in the photospheric magnetic pore regions (Dorotovic et al. 2008). Morton et al. (2010) using ROSA G-band observe a clear anti-phase correlation between the intensities and diameters of the pores. Kink waves (m = 1), are weakly compressible modes, where, the flux tube axes undergoes a periodic transverse displacement (perpendicular to the direction of the magnetic field). This mode is caused by the periodic transverse bulk motions of the plasma and therefore, periodic Doppler shifts are expected to be observed. Historically, kink waves first claimed to be observed by detecting Doppler-shift oscillations with periods of 35 to 70 seconds in H$\alpha$ spicules, using a ground-based coronaraph (Kukhianidze et al. 2006). After, kink transverse displacement of spicules were observed by (De pontieu et al. 2007) using SOT (Solar Optical

---

⋆ Contact e-mail: nedadadashi2@gmail.com
† Deceased

2  *M. Ghiasi et al.*

Telescope) aboard Hinode. Zaqarashvili et al. (2007) applied DFT (Discrete Fourier Transform) technique to the time series of the Doppler-shifts of Hα spicules and found signatures of propagating kink waves with periods in the range of 30 to 40 sec. They estimate the kink speed, and the magnetic field strength to be 90 to 110 km/s, and 12 to 15 Gauss, respectively. Transverse kink oscillations of Quiet Sun fibrils are observed and measured by a semi-automated tracking technique (Morton et al. b,a). Standing kink oscillations are observed in the coronal loops very frequently, since its first detection by Nakariakov et al. (1999); Aschwanden et al. (1999) using TRACE EUV data (Nakariakov et al. 2021). Mechanisms like resonant absorbtion and Kelvin-Helmholtz instabilities are known to be responsible for decaying these modes (Terradas et al. 010b; Su et al. 2018; Antolin & Van Doorsselaere 2019; Ruderman et al. 2019).

Morton et al. (2012) observed concurrent kink and sausage oscillations over prominence fibrils. They found periods of 135 to 241 sec, with transverse velocities of 1 to 2 km/s for the sausage mode, and 6.4 km/s velocity amplitude with periods of 3 min for the Kink mode. In this work we study the spectral properties of Fe XII 195.119 over a Crescent shaped prominence structure. Observational signatures of the all three MHD modes are investigated over this structure. Results suggest the concurrent presence of sausage and kink modes over this structure.

## 2 OBSERVATIONS AND DATA ANALYSIS

A tornado like prominence on the solar west limb is observed by EUV Imaging Spectrometer (EIS, Culhane et al. (2007)) It has a crescent like structure which looks bright in the AIA 1600 and 1700 Å channels, while, it looks dark for the other hotter passbands. The crescent structure stays quite stable over the studied time. The structure undergoes a filament eruption at the end of the observation time interval.

The EIS spectrometer aboard Hinode is a normal incidence telescope having multi-layer coatings on both its mirror and grating, which enables it to produce high-resolution image and spectra of the solar transition region and corona in the wavelength ranges of 170 to 210 Å and 250 to 290 Å. The spectral and spatial resolution of this instrument is known to be 0.0223 Å per pixel and 1 arcsec per pixel, respectively (Korendyke et al. 2006; Culhane et al. 2007). The used EIS data consists of a raster scan (with exposure time of 51 sec, and step size of 2 arcsec starting at 13:02 UTC and ending at 13:45 UTC of the 3rd April 2014 and FOV of 98″ × 256″), and a sit-and-stare sequence (with exposure time of 51 sec, slit width of 2 arcsec, starting at 13:46 UTC and ending at 14:45 UTC on the same day). EIS sit-and-stare slit position is fixed at $[x_{cen}, y_{cen}] = [986'', 43'']$ crossing the middle of the crescent structure. Using *eis_prep.pro* routine available in the SolarSoft (SSW) package, the dark current, cosmic rays, and hot pixels are removed from the EIS data and radiometric calibration is applied. The slit tilt and orbital variation corrections are performed using SSW *eis_slit_tilt.pro*, and *eis_wave_corr.pro* routines. The EIS Fe XII 195.119 Å (log[T]~ 6.20), Fe X 184.540Å (log[T] ~ 6.05), and Fe VIII 185.213 Å (log[T]~ 5.60) spectral lines are studied in this work.

The images of Atmospheric Imaging Assembly (AIA, Lemen et al. (2012)) instrument aboard the Solar Dynamics Observatory (SDO) space telescope is used with step cadence of 12 sec. An image of the Sun during the time of the observation is represented in Fig. 1. The slit position of EIS spectrometer is marked over the west solar limb. Fig. 2 demonstrates a closer look to the FOV of the studied area in different AIA channels at the beginning time of the study. The horizontal and vertical white dotted lines are plotted at fixed positions to demonstrate the location of the crescent structure. The red cuts are the artificial slits used to plot time-distance diagrams, in Figs. 12 and 13. Using *eis_auto_fit.pro* routine on the EIS spectral lines, the line intensities ($I_0$), centroid positions ($\lambda_0$), and profile widths ($\sigma$) are obtained (on each spatial pixel) by fitting a Gaussian function with a uniform background ($b$), represented by the following Equation 1:

$$I(\lambda) = I_0 \exp(\frac{-(\lambda - \lambda_0)^2}{2\sigma^2}) + b \quad (1)$$

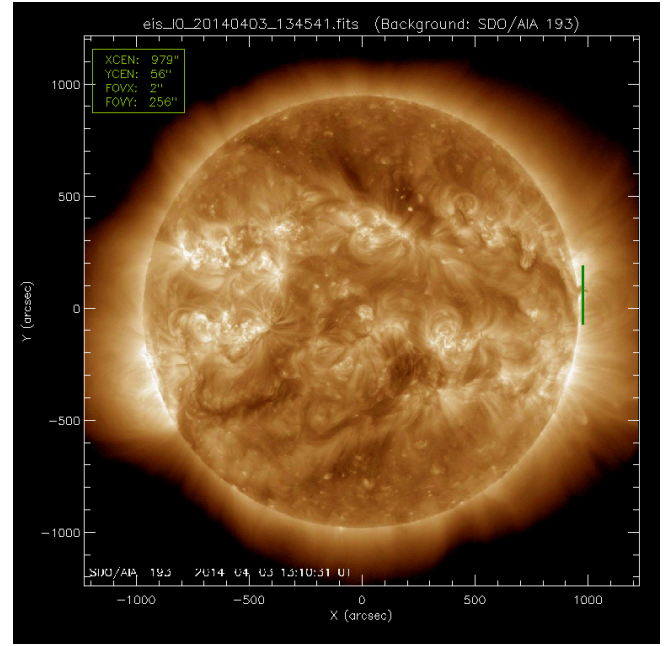

**Figure 1.** SDO/AIA 193 Å image of the sun on the day of observation. The green line shows the slit position of the EIS spectrometer.

The rest wavelength ($\lambda_{rest}$), is chosen to be the average value of the spectral line centers in the whole EIS studied field of view. Doppler shifts of the EIS spectral lines are calculated from Equation 2.

$$v = c\frac{\lambda_0 - \lambda_{rest}}{\lambda_{rest}} \quad (2)$$

## 3 RESULTS

Using the *eis_auto_fit.pro* routine on EIS spectral data the peak intensity and Doppler shift maps of Fe VII 185.213 Å, Fe X 184.540 Å, and Fe XII 195.119 Å lines are calculated and represented in Figures 3, 4, and 5, respectively. The brightest line recorded by the EIS instrument, Fe XII 195.119 Å, has a blending with Fe XII 195.179 Å line. Two Gaussian fits are used, to remove the effect of this blend. Positive (represented by red color) and negative (represented by blue color) values of Doppler-shifts represent down-flows (red shifts) and up-flows (blue shifts), respectively. Contours of zero Doppler shifts are over-plotted on the all maps. As it is visible in the intensity maps (Figs. 3 to 5), as well as Fig. 2, a prominence structure over the west solar limb is seen in all temperature channels. Above it, a crescent like structure exists that looks dimmer respect to its surroundings.

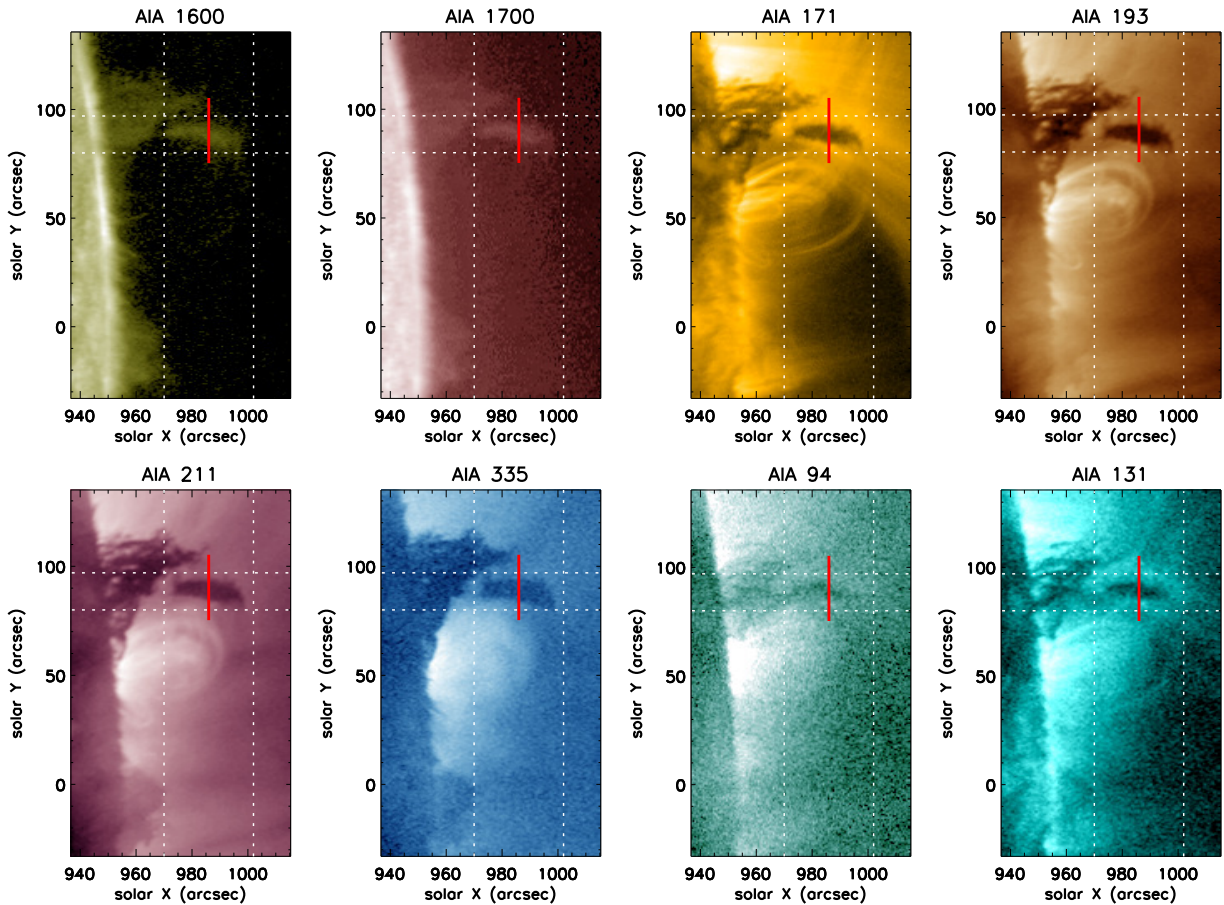

**Figure 2.** AIA/SDO images of the studied area in 1600, 1700, 171, 193, 211, 335, 94 and 131 Å channels taken on the 3rd April 2014 on 13:02:34 UTC. The horizontal and vertical white dotted lines are plotted at fixed positions to demonstrate the location of the crescent structure. The red cuts are the artificial slits used to plot time-distance diagrams (of Figs. 12 and 13). The crescent structure looks bright in AIA 1600, and 1700 cool channels, while it looks dark in the all other hotter channels.

This structure stays almost stable for about few hours and after around 14:45 UTC gets erupted.

　　Looking into the central axis of the crescent structure (in the warm coronal lines of Fe x 184.540 Å, and Fe xii 195.119 Å) one sees red shifts on one side and blue shifts on the other side with almost similar magnitudes (± 10 km/s for Fe x 184.540 Å, and ± 5 km/s for Fe xii 195.119 Å lines). Therefore, it seems that there exist some kind of rotational motions along the cental axis of the dimmed crescent structure, as well as the prominence base itself. On the other side, the crescent structure seems to have zero Doppler shifts in the cooler Fe viii 185.213 Å line. Comparison the Doppler shifts in Figs. 3 to 5), one sees that the cooler the spectral line, The stronger bulk flow in the Field of View. In the right panel of Figure 5, the line widths in the crescent cavity and the prominence base structures appear to be broader than some parts of the surroundings.

　　In Fig. 6, the three panels from left to right represent the time evolution maps of peak Intensity, Doppler shift, and the spectral line width of Fe xii 195.119 Å, respectively (using the sit-and-stare EIS data). Contours of zero Doppler shifts are over-plotted for all three panels. Intensity contours of Fe xii 195.119 Å (yellow) demonstrate the time evolution of the borders of the crescent structure.

### 3.1 Borders of the crescent structure (CS)

To obtain the position of the CS borders several ways one could select: The simplest way is to use a value of Fe xii 195.119 Å intensity for plotting intensity contours which crosses the edges of the CS structure. This method is used in Figs. 5 and 6 (yellow contours). However, before performing any detailed study about the dynamics of this structure one needs to define a more careful criterion for determining the borders of the structure. There are different methods to extract the borders in the computer vision applications, such as Sobel, Roberts, Prewitt, Laplacian, and Canny algorithms. The last one, which is the most common method for the edge detection and it looks highly effective, is used here. Canny filter using a non-maxima suppression along with applying thresholds is able to produce thinner and smoother edges, respect to the other mentioned methods (Gonzalez & Woods 1992; Alzahrani & Chen 1997; Ramnarayan et al. 2019; Lynn et al. 2021; Laigong & Sitong 2023). To do this first, we applied a gradient filter on the time evolution intensity map of Fe xii 195.119 Å (left panel of Fig. 6). This ways, the top and below borders of CS appears as two Gaussian like peaks at the two opposite sides of the CS. Then, we fitted double Gaussian functions to each column of the gradient-filtered-map to obtain the exact positions of the two top and below borders of the CS. Multiplying -1.0 to each column of the gradient-filtered-map, the cental axis position of the CS, which we defined to be the

4    *M. Ghiasi et al.*

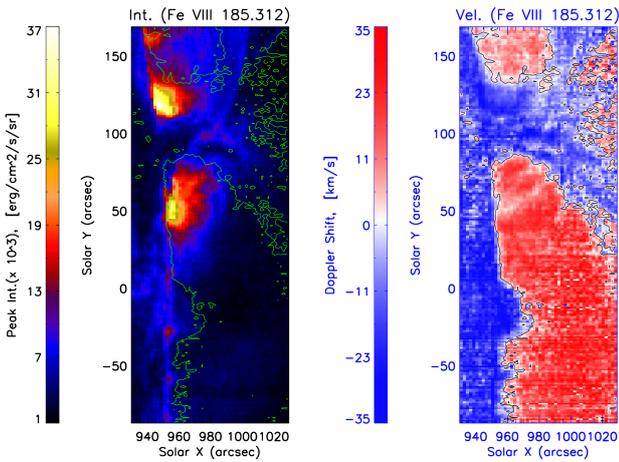

**Figure 3.** Intensity, and Doppler shift map of Fe VII 185.213 Å from the EIS raster data. Red and Blue colors represent backward and forward flows along the observation line-of-sight, respectively. green and black over-plotted contours demonstrate zero Doppler shift.

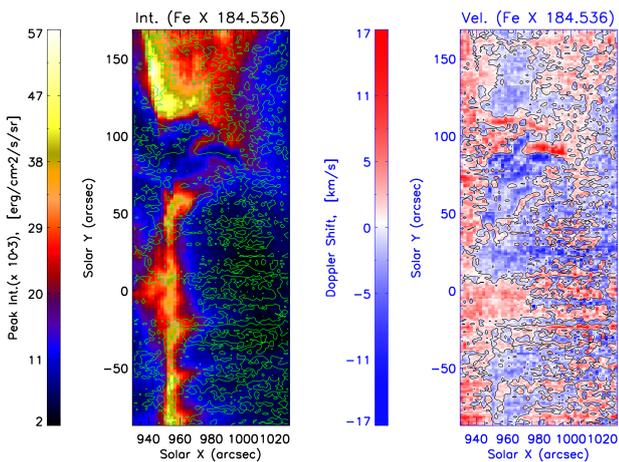

**Figure 4.** Intensity, and Doppler shift map of Fe X 184.540 Å from the EIS raster data. Green and black over-plotted contours demonstrate zero Doppler shift.

dimmest intensity pixel in each time instance in the CS, appears as a Gaussian curve. Fitting single Gaussian functions, we obtain the time evolution of the cental positions of the CS. Results are shown in Fig. 7. Blue, yellow, and red contours on the second panel from the top left show the time evolution of the positions of top border, central axis, and below border of the CS. Top panel of the right column shows the time variation positions of the top border (blue), central axis(black), and below borders (red). Using the IDL least-square curve fitting routine, CURVEFIT, sinusoidal curve fits with following equation

$$y(t) = y_0 + y_{00} \sin(\omega t + \phi_0) \quad (3)$$

are applied to the borders (solid thin blue and red lines) and central axis (purple dot-dashed line.) The last one, with some positive and negative y-offset are over-plotted to the top and below borders, for comparison. In the Equ. 3, the $y_0$ is the height of the bottom of the CS in the Solar-Y direction, $y_{00}$, and $\omega$, and $\phi_0$ are the amplitude, radial frequency, and initial phase of the CS borders' transverse

Oscillations. The coefficients of the fit are summarized in Table 1. The middle Panel of the right column shows the time evolution of the thickness of the SC (by subtracting the top border positions from below ones). Results show the thickness increase with time and undergoes some oscillations. Therefore, a sinusoidal function with a linear increasing background, as following

$$D(t) = D_0 + V_0 t + D_{00} \sin(\omega_0 t + \phi_{00}) \quad (4)$$

is fitted to the thickness of the CS (pink dot-dashed line). where $D_0$ is the minimum thickness of the SC, $v_0$ is the expanding rate of the cross section, $D_{00}$, and $\omega_0$, and $\phi_{00}$ are the amplitude, radial frequency, and initial phase of the SC's thickness Oscillations. The coefficients of the fit are summarized in Table 2. The below Panel of the right column shows the asymmetry of the top-half and below-half thickness of the CS with respect the obtained central axis. In almost whole time interval of this study the top-half thickness of CS is larger except in the short time intervals of (0 to 4 min), (43 to 47 min), (49 to 50 min), and (52 to 55 min).

Left panel of Fig. 8, from top to below demonstrates a closer look into the SC from Fe XII 195.119 Å line's EIS raster peak intensity, EIS sit-and-stare peak intensity, EIS sit-and-stare Doppler shifts, and EIS sit-and-stare spectral line width, respectively. The intensity, Doppler shift, and the line width scales are the same as those of the Fig. 6. The white contours represent the time evolution of the top and below border of the CS. Light green contours represent the zero Doppler shifts. The two top right panels are similar to those of the Fig. 7, (added to make the comparison easier.). The pink and blue arrows denote the occurring time instances of the CS'S maximum and minimum thickness. The linear blue line shows the background of the fitted function to the CS thickness. The third panel from top of the right column, shows the Doppler shifts of the Fe XII 195.119 Å line at top border, Central axis, and below border of the CS structure. A sinusoidal fit with constant background, as following

$$v(t) = v_0 + v_{00} \sin(\omega_D t + \phi_{0D}) \quad (5)$$

is applied on each of the three mentioned time series (Equ. 5). where $v_0$ is the background flow Doppler shift, $v_{00}$, $\omega_D$, and $\phi_{0D}$ are the amplitude, radial frequency, and initial phase of the Doppler shift Oscillations. Result of the Doppler shift fits are over-plotted by solid blue, green, and red lines. The fit coefficients are summarized in the Table 3. The lowest panel on the right column shows the Fe XII 195.119 Å spectral width of Fe XII 195.119 Å line over the CS's top border (blue), central axis (green), and below border (red), in km/s unit. It seems that they are uncorrelated in most of the time and show oscillatory behavior. More detailed analysis, provided in the 'Fe XII 195.119 Å line width over the CS' subsection, in continue.

### 3.2 Kink wave in the CS

As mentioned, kink waves are the periodic bulk motions of plasma perpendicular to the direction of the magnetic field. Therefore, Doppler shift Oscillations are expected to be observed. As shown in right top panel of Figures 7, and 8, transverse movements (perpendicular to the CS axis) are observable for the all top border, central axis, and below border, of the CS. To measure it quantitatively, sinusoidal functions are fitted to the time series of the borders. Since each EIS pixel is 1″, considering a 720 km per 1″ ratio, the top border undergoes a transverse oscillation, as: $y_{top}(t) = (1130\,\text{km})\sin(0.001570\,t + 216°)$ with period of $66.7 \pm 18.8$ min. The central axis, and below border show oscillations in the Solar Y direction, as, $y_{center}(t) =$

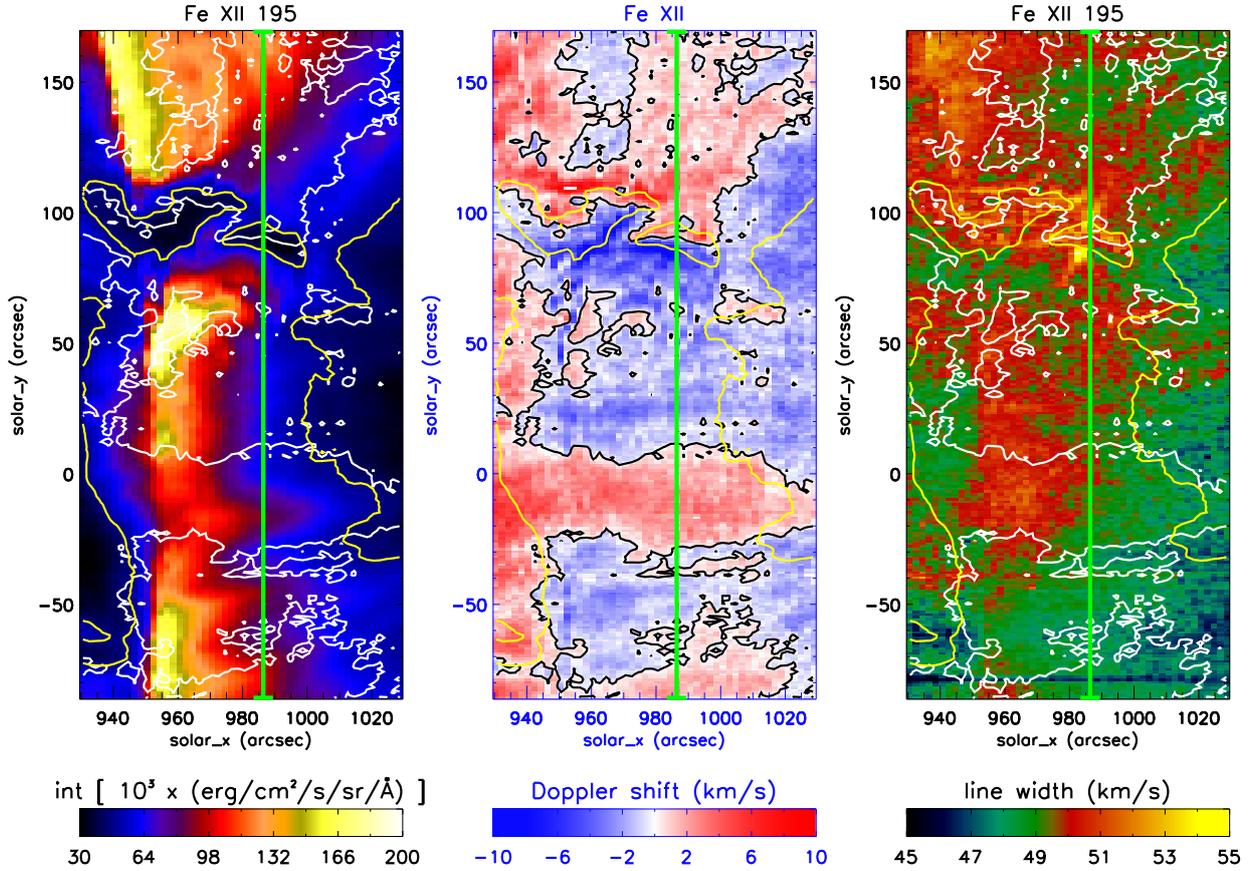

**Figure 5.** Left, middle, and right panels represent the Intensity, Doppler shift, and line width maps of Fe XII 195.119 Å from EIS raster data. Contours of Zero Doppler shift are over-plotted on the all three panels. Intensity contours of Fe XII 195.119 Å (yellow) demonstrate the borders of the crescent structure. Light green lines show the position of the sit-and-stare sequence slit.

**Table 1.** Coefficients of the sinusoidal curve fitted to the borders of the crescent structure $[y(t) = y_0 + y_{00}\sin(\omega t + \phi_0)]$.

|  | $y_0$ (pix) | $y_{00}$ (pix) | $\omega$ (mHz) | Period (min) | $\phi_0$ (degree) |
|---|---|---|---|---|---|
| top border | 179.3 | 1.57 | 1.570 | 66.7 | 216 |
| (error) | (±0.5) | (±0.61) | (±0.444) | (±18.8) | (±55) |
| central axis | 174.2 | 1.93 | 1.818 | 57.6 | 164 |
| (error) | (±0.4) | (±0.64) | (±0.292) | (±9.3) | (±33) |
| below border | 169.7 | 1.10 | 2.101 | 49.9 | 143 |
| (error) | (±0.4) | (±0.65) | (±0.458) | (±10.9) | (±55) |

**Table 2.** Coefficients of the sinusoidal curve fitted to the width of the crescent structure $[D(t) = D_0 + V_0 t + D_{00}\sin(\omega_0 t + \phi_{00})]$.

|  | $D_0$ (pix) | $V_0$ (pix/s) | $D_{00}$ (pix) | $\omega_0$ (mHz) | Period (min) | $\phi_{00}$ (degree) |
|---|---|---|---|---|---|---|
| value | 8.41 | 0.03614 | 1.0000 | 5.030 | 20.8 | 339 |
| (error) | (±0.32) | (±0.0086) | (±0.0001) | (±0.218) | (±0.9) | (±24) |

**Table 3.** Coefficients of the sinusoidal curve fitted to the Doppler shifts over the borders of the crescent structure $[v(t) = v_0 + v_{00}\sin(\omega_D t + \phi_{0D})]$.

|  | $v_0$ (km/s) | $v_{00}$ (km/s) | $\omega_D$ (mHz) | Period (min) | $\phi_{0D}$ (degrees) |
|---|---|---|---|---|---|
| top border | 0.13 | 3.4 | 2.223 | 47.1 | 230 |
| (error) | (±0.17) | (±0.2) | (±0.081) | (±1.7) | (±9) |
| central axis | -4.04 | 3.2 | 2.148 | 48.8 | 256 |
| (error) | (±0.18) | (±0.2) | (±0.103) | (±2.3) | (±11) |
| below border | -7.10 | 2.4 | 2.354 | 44.5 | 242 |
| (error) | (±0.20) | (±0.3) | (±0.121) | (±2.3) | (±15) |

(1390 km) sin(0.001818 t + 164°) with period of 57.6 ± 9.3 min, and $y_{below}(t) = (792 \text{ km})\sin(0.002101 t + 143°)$ with period of 49.9 ± 10.9 min, respectively. Considering the error ranges, one could interpret the initial phases, rather in-phase. Over-plotting the $y_{center}(t)$ fitted function to the top and below borders, by adding up some offsets (purple dot-dashed lines in the mentioned Figures 7, and 8), shows a quite in-phase transverse movement of the CS, which could be an indication of the kink waves, with an oscillatory transverse motion in the Solar-Y direction. The transversal oscillat-

6   *M. Ghiasi et al.*

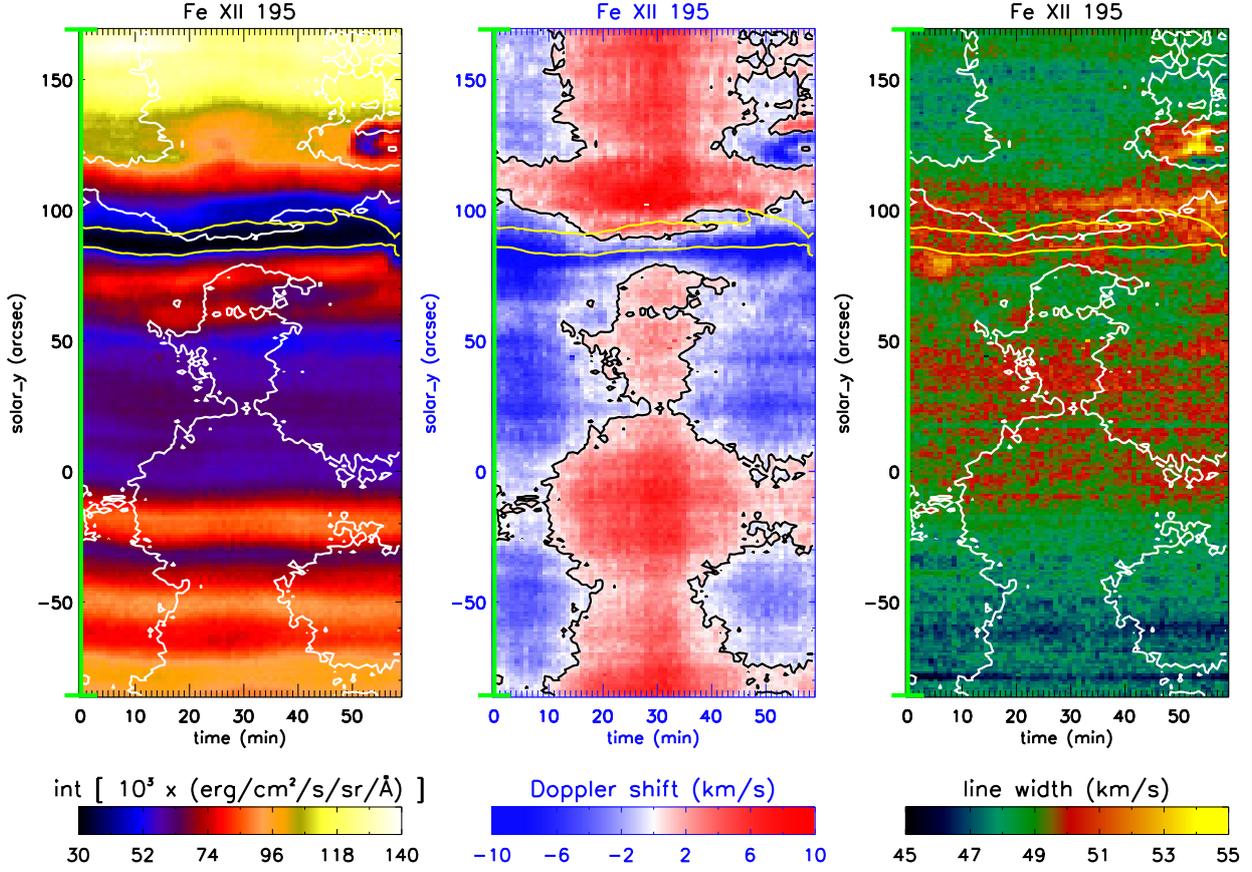

**Figure 6.** From Left to right; **First panel** represents the peak intensity map of Fe XII 195.119 Å from the sit-and-stare data obtained by EIS. The over-plotted light green line shows the slit position. Second, third panels demonstrate the time evolution maps of Doppler shift, and spectral line width, respectively. Red and Blue colors represent downflows and upflows along the observation line-of-sight, respectively. Contours of zero Doppler shifts are over-plotted for the all three panels. Intensity contours of Fe XII 195.119 Å (with yellow color) demonstrate the time evolution of the borders of the crescent structure.

ing velocities are obtained to be as,

$$v_{\text{top border}} = \frac{2 \times \pi \xi_{\text{top}}}{\text{kink period (top border)}} = 1.77 \, \text{km/s}, \quad (6)$$

$$v_{\text{central axis}} = \frac{2 \times \pi \xi_{\text{center}}}{\text{kink period (central axis)}} = 1.53 \, \text{km/s}, \quad (7)$$

$$v_{\text{below border}} = \frac{2 \times \pi \xi_{\text{below}}}{\text{kink period (below border)}} = 1.66 \, \text{km/s}, \quad (8)$$

where, the $\xi$ is the maximum amplitude of the transverse displacement. However, Some of the data points (especially on the top border, between time 20 to 40 min) seem to deviate from the purple dot-dashed fitted line. In the next subsection we try to investigate a possible reason for this behaviour.

In continue, we studied the Doppler shift variations of Fe XII 195.119 Å line over the all borders of the CS structure (third panels from top, Fig. 8). Red, and blue colors show the directions of the bulk motions to be against, and along the LOS (Line-Of-Sight), respectively. Between 17 to 41 minutes, it seems that a rotational motion exists around the central axis of the CS, with an average velocity of about 5 km/s. Considering a minimum of 4.5 arcsec radius for the CS cross section, along with a 720 km per 1″ ratio, a centripetal accelerations of $a = \frac{v_{oo}^2}{r} = 7.7 \, \text{m/s}^2$ is obtained, which is quite low. Beside that, a universal oscillatory pattern in the Doppler shift is visible over the all borders. Top border clearly experience an alternative red and blue Doppler shifts. However, changing the sign of red/blue shifts do not occur for central axis and below borders. We should recall that the used rest wavelength to calculate the Doppler shifts is not absolute, and it is obtained by averaging the central positions of Fe XII 195.119 Å line over the all study time interval (due to the lack of calibration lamp on EIS). Therefore, the real value of the rest wavelength might be somewhat different, which could end up in changing the direction of the plasma bulk motion. For instance, if we switch up the rest wavelength for about (3 km/s), the central axis also experience the sign change for the Doppler shift.

Applying sine fits to the time series of the Doppler shifts over the all borders, we obtained:
$v_{\text{top}}(t) = 0.13 + 3.4\sin(0.002223\,t + 230°)$ with period of $47.1 \pm 1.7$ min, $v_{\text{center}}(t) = -4.04 + 3.2\sin(0.002148\,t + 256°)$ with period of $48.8 \pm 2.3$ min, and $v_{\text{below}}(t) = -7.10 + 2.4\sin(0.002354\,t + 242°)$ with period of $44.5 \pm 2.3$ min, respectively. The Obtained initial phases on top, and below border, and the central axis of the CS seems to be in phase (considering their error ranges). This would suggest the existence of kink mode with an oscillatory transverse motion in the observers LOS (line of sight).

One explanation for observing kink modes oscillating in two

*Concurrent Kink and Sausage Waves* 7

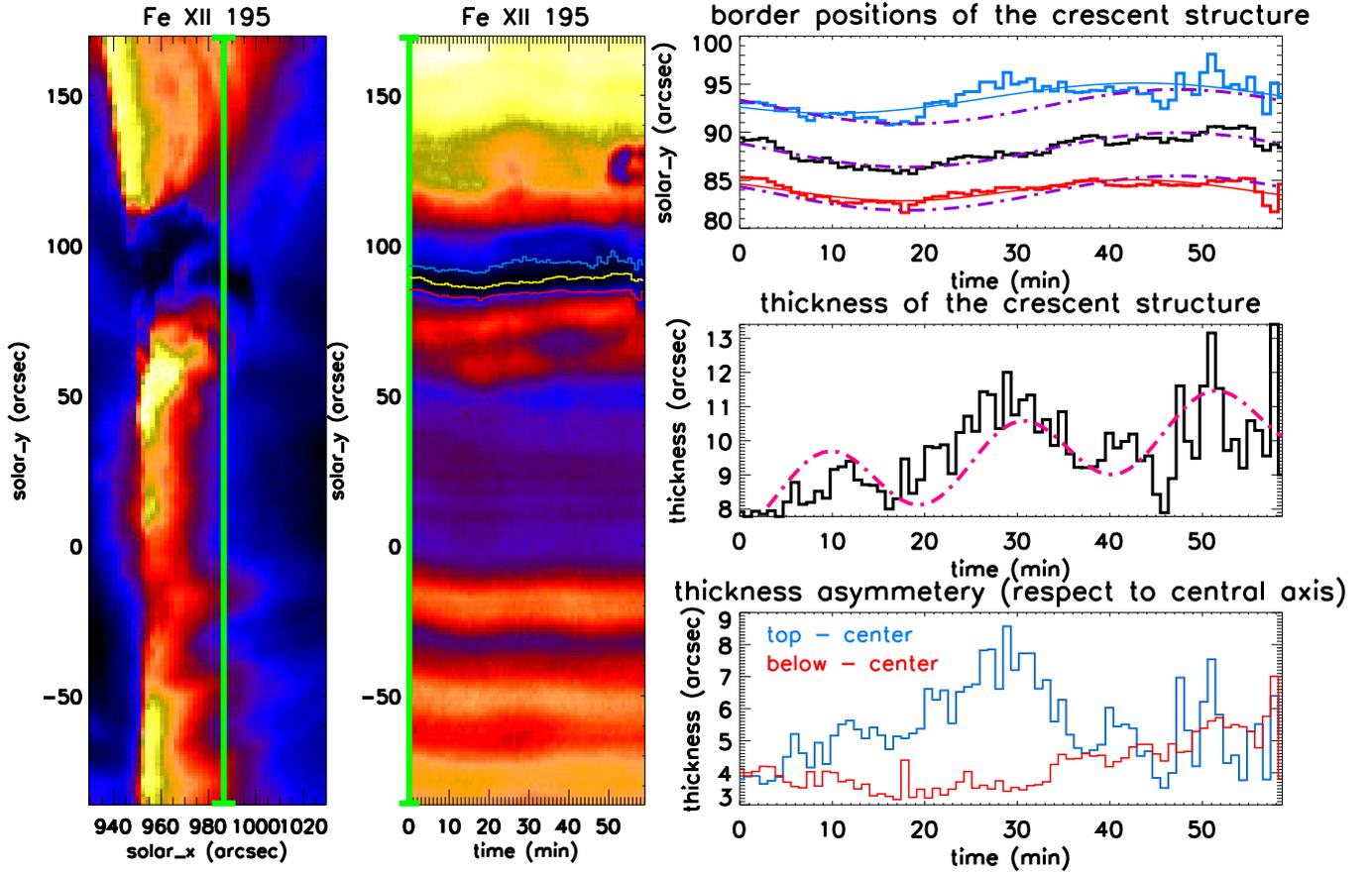

**Figure 7.** Left panel: Fe XII 195.119 Å intensity map from the EIS raster data. The light green line shows the position of the EIS slit for sit-and-stare sequence data. Second panel from left, shows the Fe XII 195.119 Å intensity map from the EIS sit-and-stare data. Blue, yellow, and red contours on this panel show the time evolution positions of top border, central axis, and below border of the CS obtained by gradient method. Top panel of the right column shows the time variation positions of the top border (blue), central axis(black), and below borders (red). Sinusoidal curve fits with equation [y(t) = y$_0$ + y$_{00}$sin($\omega$t + $\phi_0$)] are applied to the borders (solid thin blue and red lines) and central axis (purple dot-dashed line.) The last one, with some positive and negative y-offset are over-plotted to the top and below borders, for comparison. The middle Panel of the right column shows the time evolution of the thickness of the SC. A sinusoidal function [D(t) = D$_0$ + V$_0$t + D$_{00}$sin($\omega_0$t + $\phi_{00}$)] is fitted to the thickness of the CS (pink dot-dashed line). The below Panel of the right column shows the asymmetry of the top-half and below-half thickness of the CS with respect the obtained central axis.

different perpendicular directions (in the Solar-Y, and in the LOS directions), could be that we are observing the projections of the kink transverse oscillations, in two different perpendicular directions. Fig. 9 shows the schematic of the both kink modes observed in our study. Considering the main kink oscillatory direction having an angle of $\theta$ with respect to the observer LOS, and assuming an oscillatory equation for that (with a constant velocity amplitude of $v_{0T}$, radial frequency of $\omega_T$, and initial phase of $\phi_{0T}$) as following,

$$v_T(t) = v_{0T} \sin(\omega_T t + \phi_{0T}) \quad (9)$$

the tangential components of the main kink wave should be observable in the Solar-Y direction as,

$$y'(t) = y'_{00} \sin(\omega t + \phi_0) = \sin(\theta) \, v_{0T} \sin(\omega_T t + \phi_{0T}). \quad (10)$$

Therefore, along the LOS we expect to see

$$v(t) = v_{00} \sin(\omega_D t + \phi_{0D}) = \cos(\theta) \, v_{0T} \sin(\omega_T t + \phi_{0T}). \quad (11)$$

Solving these equations simultaneously, along with considering the time derivatives of the Equs. 10, and 11 to be continuous, we have obtained the properties of the main kink oscillation, for instance, for the top border to be as following:

- $\theta = 27°$,
- $v_{0T} = 5.32$ km/s,
- $\omega_D = 0.0057071$ Hz,
- period = 33.44 min,
- $\phi_{0T} = 0.5823$ rad.

Fig. 5 of the Antolin et al. (2018) represent a couple of schematic Doppler Shift maps, as a result of a 3D MHD simulation of spicule like structures. The Doppler shift pattern obtained from the case of rotation coupled with the kink mode looks to be similar to our findings represented in the Doppler shift map of the Fig. 8. Supporting the idea of coupling rotation with the kink mode, which we provided observational signatures for both of them.

8 *M. Ghiasi et al.*

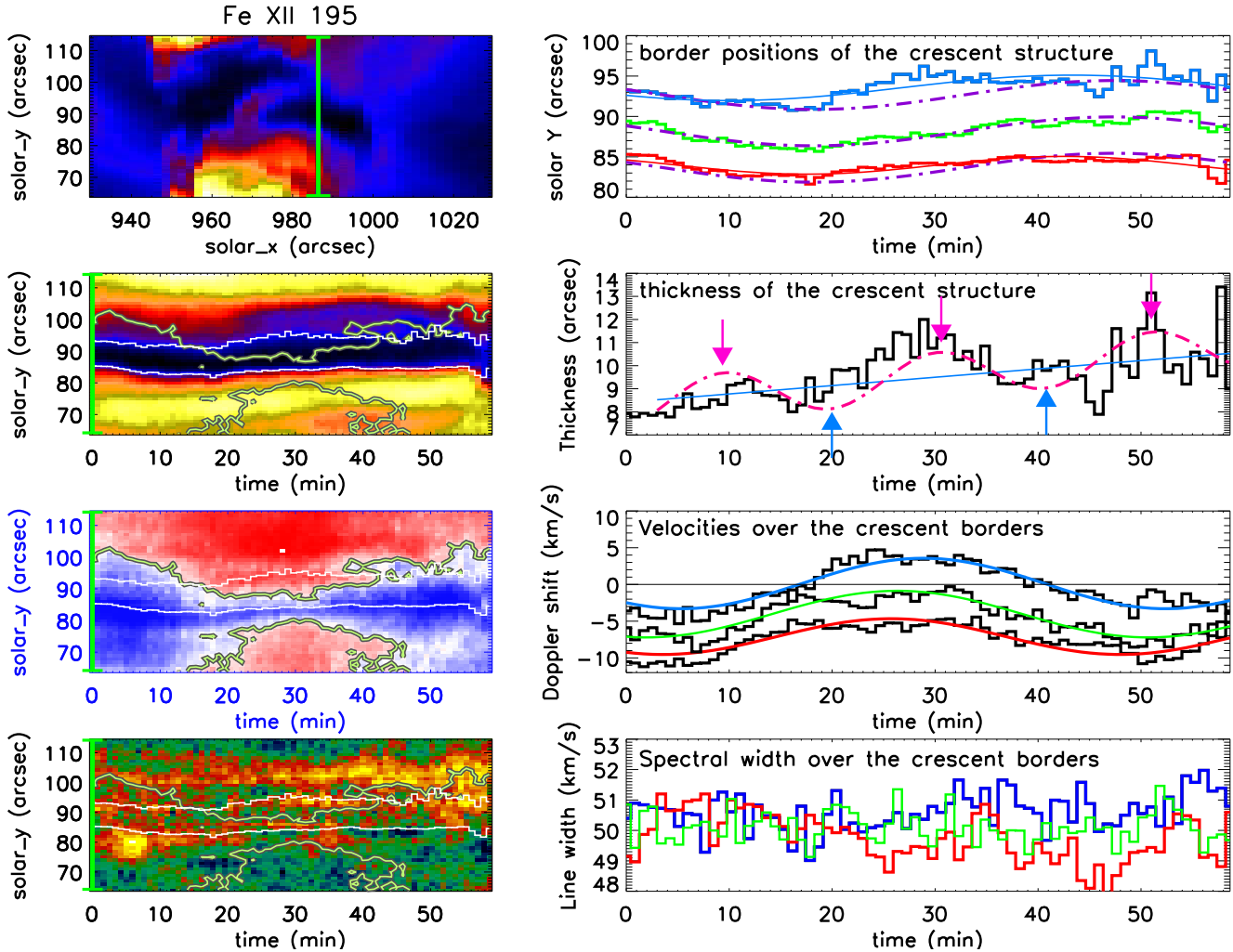

**Figure 8.** Left panel, from top to below demonstrates a closer look into the SC from Fe XII 195.119 Å line's EIS raster peak intensity, EIS sit-and-stare peak intensity, EIS sit-and-stare Doppler shifts, and EIS sit-and-stare spectral line width, respectively. The white contours represent the time evolution of the top and below border of the CS. Light green contours represent the zero Doppler shifts. The two top right panels are similar to those of the Fig. 7, (to better comparison). The pink and blue arrows denote the occurring time instances of the CS'S maximum and minimum thickness. The third panel from top of the right column, shows the Doppler shifts of the Fe XII 195.119 Å line at top border, Central axis, and below border of the CS. A sinusoidal fit with constant background, as $v(t) = v_0 + v_{00}\sin(\omega_D t + \phi_{0D})$, is applied on each of the three mentioned time series. Result of the Doppler shift fits are over-plotted by solid blue, green, and red lines. The lowest panel on the right column shows the Fe XII 195.119 Å spectral width over the CS's top border (blue), central axis(green), and below (red) borders (in km/s).

### 3.3 Sausage wave in the CS

Top panel of Fig. 10 shows the Fe XII 195.119 Å peak intensity time variations. Contours of top and below borders of the CS represented with white, while, central axis is shown with yellow. Below panel demonstrates the thickness of the CS (in km) with its errors (blue line), along with the intensity variation inside the dark crescent structure with its errors (red line). The dotted vertical lines show the positions of the CS's minimum thickness according to the overplotted sinusoidal fit to the CS thickness, as mentioned before (pink curve). The anti-phase relation between the intensity and thickness observable in this Figure suggests the presence of the sausage mode in the CS (Morton et al. 2010, 2012; Verth & Jess 2015; Gomez et al. 2020). The coefficients of the sinusoidal fit (Equ. 4) to the CS thickness indicate that this sausage wave have a period of 20.8 ± 0.9 minutes, with maximum transverse velocity ($\frac{2\pi D_{00}}{\text{Period}}$) of 3.6 ± 0.2 km/s. The exceeding points from the top border (as mentioned in the last subsection) that are visible in the two top right panels of Figs. 7, and 8 between the 20 to 40 minutes, might be the effect of the combination of kink and sausage modes, which we suggest that are occurring simultaneously in the studied CS structure.

To ensure that the obtained wave (with a period of 20.8 min) categorizes as a sausage mode, we checked the anti-phase relationship between the intensity and thickness of the CS through a Fourier phase lag analysis (the section 2.2.5 of Jess et al. (2023)). If the CS intensity and thickness time series denote with (INT) and (THICK), and F and $\bar{F}$ demonstrate the Fast Fourier Transform (FFT) and its complex conjugate, respectively, the Fourier cross-power spectrum ($\Gamma_{1,2}(\nu)$) between the FFTs of INT and THICK can be calculated from the Equ. 12. The cross-power spectrum is a complex array ($\Gamma_{1,2}(\nu) = d(\nu) + i\,c(\nu)$), where, $d(\nu)$ and $c(\nu)$ are the real (co-spectrum) and imaginary (quadrature spectrum) parts of the cross-power spectrum, respectively (Bendat & Piersol (2000)).

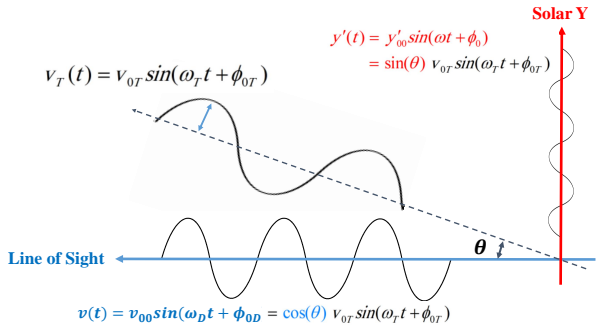

**Figure 9.** Schematic of the superposition of the two kink modes observed in the two perpendicular directions of Solar-Y, and LOS, along with the main kink mode.

$$\Gamma_{1,2}(\nu) = F(INT) * \overline{F(THICK)}. \quad (12)$$

The phase lag between the CS intensity (INT), and thickness (THICK) time series is actually the phase of the cross-power spectrum, which generally could be in the range of ± 180 degrees, and could be obtained as Equ. 13,

$$\Phi(\nu) = \arctan(\frac{<c(\nu)>}{<d(\nu)>}). \quad (13)$$

Fig. 11 show the Fourier power spectrums of the INT and THICK time series along with the real part of the Fourier cross-power spectrum, and the phase lag diagram. The horizontal axes demonstrate the frequency in the units of [1/min]. The vertical solid blue lines show the corresponding frequency of the dominant period T = 20.8 min, which is $\nu$ = 0.0481 min$^{-1}$ = 0.802 mHz. The vertical dashed lines show the error limits in frequency. Since the time series only span over an almost three periods, FFT would not be a very trusting analysis. However, the power spectrums of INT and THICK show peaks nearby the dominant period (T = 20.8 min). Moreover, the co-spectrum shows a peak exactly at this period. The corresponding phase lag at this period (20.8 min) obtained to be -178.1 degrees, which means the CS intensity is -178.1 degrees lagging behind the CS thickness, verifying the identification of the Sausage mode in the CS.

Time-distance diagram over the red cut (crossing the CS, as shown in Fig. 2) for each passbands of AIA are provided in Fig. 12. Red over-plotted lines on each panel are the intensity contours of AIA 193 Å channel which has most of its radiation from the Fe XII 193 line (with similar formation temperature as of the EIS Fe XII 195 line), and selected in a way to outline the top and below borders of the CS structure. The thickness of the CS looks to decrease as going to hotter passbands like AIA 94, and 131. A small jet like event occurring over the below border of the CS, around the 20 minutes (with positive slopes of about +19.2 km/s) in AIA 171, and 131 channels (and very faintly visible in 94 channel). Slightly after, a small jet with a negative slope (of about −4.8 km/s) emerges at the same position in AIA 193 channel.

Fig. 13 demonstrates the corresponding running difference diagrams of the Fig. 12, with dt = 240 seconds. No image enhancement technique is used. A few pulses of brightening, following by dimming are visible in almost all channels. The two strong dark pulses (starting around 20 and 40 minutes) are correlated with the minimum thickness of the CS (shown by blue arrows in Fig. 8). There-fore, occurring these jets and their afterward dimming process, might be responsible for triggering the observed sausage mode.

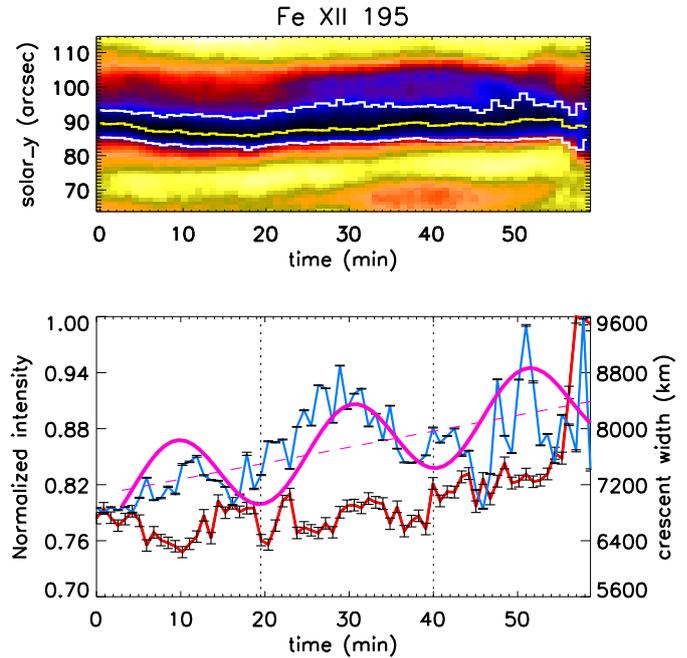

**Figure 10.** Top panel shows the Fe XII 195.119 Å peak intensity time variations. Contours of top and below borders of the CS represented with white, while, central axis is shown with yellow. Below panel demonstrates the thickness of the CS (in km) with its errors (blue line), along with the intensity variation inside the dark crescent structure with its errors (red line). The dotted vertical lines show the positions of the CS's minimum thickness according to the over-plotted sinusoidal fit to the CS thickness, as mentioned before (pink curve). The oscillations of the CS thickness and the CS intensity seems to be anti-correlated which is an strong indicators of the presence of sausage mode.

### 3.4 Fe XII 195.119 Å line width over the CS

Zaqarashvili (2003) and Erdélyi & Fedun (2007) stated that torsional Alfvén waves propagating in cylindrical flux tubes could create simultaneous red-shift and blue-shift that could make a systematic non-thermal spectral line broadening. The signature of this periodic line broadening should be observable in the FWHM of the spectral lines. Jess et al. (2009) have observed this phenomena over their studied bright point area and interpret that as signature of the presence of torsional Alfvén waves in the lower atmosphere. They found the oscillatory periods to be in the range of 2.1 to 12 minutes.

To analyse the oscillatory pattern seems to be existed in the line widths of the Fe XII line over the CS border, a Morlet 6 wavelet analysis (with the method and code provided by Torrence & Compo (1998)) is applied to the line width time series over the all borders of the CS (Figs. 14 to 16). Panel a, in all these three Figures shows the time series of the Fe XII 195.119 Å line widths (in km/s) over the analysed border. Panel b, and d represent the wavelet power spectrum, and global wavelet spectrum (summed over time), respectively. Panel c demonstrates the histogram of the wavelet phase distribution. The light green contours outline the 95% confidence level. Area below the white solid line is the cone of influence (COI).

10   *M. Ghiasi et al.*

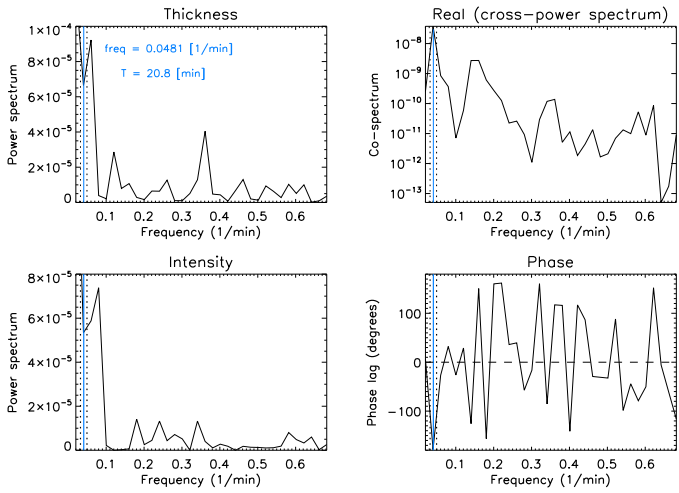

**Figure 11.** Left panels show the Fourier power spectrums of the CS thickness and intensity time series. Right panels represent the real part of the Fourier cross-power spectrum along with the phase lag diagram. The horizontal axes demonstrate the frequency in the units of [1/min]. The vertical solid blue lines show the corresponding frequency of the dominant period T = 20.8 min, which is $\nu = 0.0481$ min$^{-1}$ = 0.802 mHz. The vertical dashed lines show the error limits in frequency. The results show that the phase lag between the CS intensity and thickness to be -178.1 degrees at the dominant period (20.8 min), which verifying the anti-phase relation between these two quantities.

As the global wavelet spectrum shows, the Fe XII line width over the below border represent a weak 7.7 min oscillation. The peak of the histogram of these oscillations falls around −95°. The central border of the CS shows peak in the global wavelet diagram over the 9.9 and 6.4 minutes. Corresponding phase histograms peaks on −80° and +65°. Fe XII line widths over the top border show 2.3 , 5.9, and 10.8 min oscillations, with three different peaks in the phase diagram as, −160°, −80°, and +65°. Two of the phase angle peaks are similar over the top border and central axis. If the torsional Alfvén waves would have been presented in the CS, we would get a similar oscillation period over the top and below borders, with 180 degrees of phase difference. These, results do not show such oscillatory patterns.

To investigate the radial variations of Fe XII 195.119 Å spectral line properties over the CS cross section, the line peak intensity (left column), Doppler shift (middle column), and line width (right column) are plotted as a function of structure's height (Solar-Y ) across the CS, at moments represented by blue and pink arrows in the second top right panel of Fig. 8, in the Figs. 17 and 18. Red, green, and blue solid vertical lines denote the positions of the top border, central axis, and below border of the CS, at each studied frame time.

Radially getting distance from the CS's central axis, the intensity and amount of the Doppler shifts clearly increase, reminding of the radial velocity diagram of a solid rotating object. This probably means that, inside the CS should have much higher density respect to its surroundings. Though the CS looks dark in the hot EUV passbands (Fig. 2), it looks bright in AIA 1600 and 1700 channels, which may suggest that the CS should be filled with the cooler materials. Therefore, we need to have an estimation of the density of the CS.

Levens et al. (2015) using the 7.1.3 version of CHIANTI atomic database (Dere et al. 1997; Landi et al. 2012), with line-ratio approach, have calculated the electron density of a tornado-like prominence structure to be equal to $\log(n_e)$ = 8.5 for Fe XII 195.119 Å line (with formation temperature of T = 1.6 MK). They have used the density sensitive doublet lines of Fe XII 195.119 Å and 195.179 Å (Fig. 19).

Using the same technique the average line ratio in the CS to be obtained as 26.5, which indicates electron densities of $\log(n_e)$ = 9.3 (Fig. 20).

Looking to the right panels of Figs. 17 and 18, one seems that the line widths of the central axis experience a local minimum (except over the minutes 20.0), while, moving from central axis toward the CS borders, the line widths increase. However, Levens et al. (2015) reported increasing non-thermal line width inside their observing tornado. Though the line width errors in our study are a bit large (respect to their changing range), still we can not say an apparent line width increase is observed in the CS, which exclude the probability of the prepense of torsional Alfvén waves in the CS.

### 3.5   the CS cross section's increasing rate

Another evident point from the result of sinusoidal fit to the CS thickness is that the CS cross section increase with constant velocity of $V_0 = 26.0 \pm 0.2$ km/s as time move forward. Considering the magnetic flux conservation equation ($A_{t_1}.B_{t_1} = A_{t_2}.B_{t_2}$, $A_{t_1,t_2} = \pi D_{t_1,t_2}^2$), where $A_{t_1}$ and $B_{t_1}$ are the cross section and magnetic field strength at the time instance $t_1$, and $D_{t_1}$ is the radius of the cross section or the CS thickness, indicate that the magnetic field strength in the CS should decrease with increasing time. If we consider the $t_1$, and $t_2$ to be the starting and ending time instances of our EIS sit-and-stare observation, the rate of changing magnetic field would become as following:

$$\frac{B_{t_2}}{B_{t_1}} = \left(\frac{D_{t_1}}{D_{t_2}}\right)^2 = \left(\frac{8.41 \text{ pixels}}{10.80 \text{ pixels}}\right)^2 = 0.61 \quad (14)$$

Therefore,

$$\frac{\Delta B}{B_{t_1}} = \frac{B_{t_2} - B_{t_1}}{B_{t_1}} = \frac{(0.61 - 1)B_{t_1}}{B_{t_1}} = -0.39 \quad (15)$$

$$\frac{dB}{dt} = \frac{-0.39 \, B_{t_1}}{67.0 * 51.0 \text{ sec}} = -1.14 \times 10^{-4} \, B_{t_1} \text{ [per sec]} \quad (16)$$

### 4   CONCLUSIONS

In this study, we measured the Intensity, Doppler shift, and line width of coronal Fe XII 195.119 Å line over a crescent structure (CS) in a limb solar prominence. Borders and central axis positions of the CS are obtained with edge detection Canny technique. We conclude our results as following,

• Doppler maps of the studied spectral lines represent a clockwise rotation along the central axis of the CS (Figs. 5). The rotational flows seem to be two times stronger in the Fe X 184.540 Å line (±10 km/s), respect to Fe XII 195.119 Å line (±5 km/s). While, the surrounding areas of the CS in Fe VIII 185.213 Å line, show much stronger flows, inside the CS seems to have zero Doppler shift in this line (Fig. 3).

• Transverse kink oscillations are observed both in the Solar-Y direction and in the Doppler shifts over the observers' LOS (Fig. 8). One explanation could be the oscillatory direction of the main kink wave have an angle with the observers LOS. This angle is calculated to be 27 degrees for the CS top border. The main kink amplitude velocity and periods are obtained to be 5.3 km/s, and 33.4 minutes, respectively.

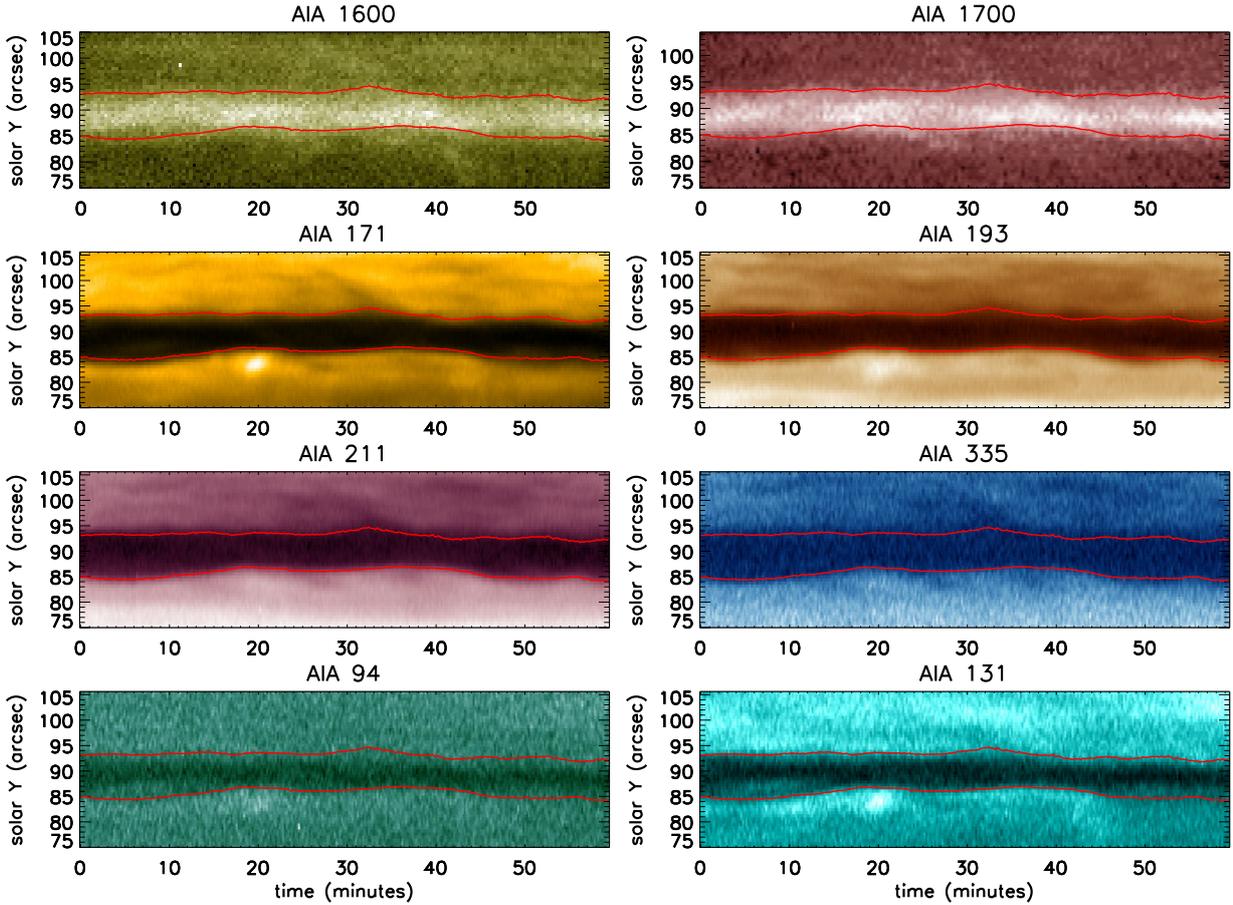

**Figure 12.** Time-distance diagram over the red cut (crossing the CS, as shown in Fig. 2) for different passbands of AIA. Red over-plotted lines on each panel are the intensity contours of AIA 193 Å channel, and selected in a way to outline the top and below borders of the CS structure.

• The thickness of the CS is studied and found to increase, Periodically. A sinusoidal fit to the CS thickness shows a dominant period of 20.8 minutes. Fourier phase lag analysis suggests the time series of the CS intensity lags the CS thickness with an angle of -178.1 degrees. Anti-correlation of the brightness of the CS and its thickness suggests the presence of the sausage modes with a prevalent period of 20.8 minutes. AIA time-distance plots show the existence of localized tiny plasma jets over the below border of the CS, and the dimming pulses looks to occur when the CS has its minimum thickness. Therefore, it is suggested that the jets and their afterward dimming are responsible to trigger the sausage mode.

• Morlet wavelet analysis with 95% confidence level is used to investigate the oscillatory pattern of the Fe XII 195 Å line, over the CS borders and central axis. Results show oscillations in the range of 2.3 to 10.8 minutes with two main common phases of −80°, and +65° (Figs. 14 to 16). If the torsional Alfvén waves would have been presented in the CS, we would get a similar oscillation period over the top and below borders, with 180 degrees of phase difference. These, results do not show such oscillatory patterns.

• The central axis of the CS seems to have the lowest line width at the studied time instances. moving from center to limb of the cross section of the CS, an slight increase in the Fe XII 195 Å line width is observed (Figs. 17 and 18), which is in contrary with our expectations for the case of Torsional Alfvén waves (Levens et al. 2015).

• The CS rotates along its central axis (between the time of 20 and 40 minutes), similar to a rigid body, which means, moving in the radial direction in the CS cross section, the amount of the Doppler shift increases from zero to about 6 km/s (Figs. 17 and 18). Therefore, inside the dark CS should have much higher density respect to its surroundings.

• Using the density sensitive pair of Fe XII 195.119 Å and 195.179 Å lines, the average of the electron densities over the CS are obtained to be log($n_e$) = 9.3 (Fig. 20).

• Considering the magnetic flux conservation inside the CS, the constant expanding background of the CS thickness, causes the magnetic field to decay with the rate of $\frac{dB}{dt} = -1.14 \times 10^{-4}$ $B_{t_1}$ per second, where $B_{t_1}$ is the initial magnetic field.

• The sound speed in the coronal Fe XII 195.119 Å line is calculated to be equal to $C_s = \sqrt{\frac{k_B T}{m_{Fe\ XII\ ion}}}$ = 15.3 km/s. Considering the length of the CS to be about L = 30 arcsec, the kink speed $C_k$ is obtained from the Equ. 7.2.4 of Aschwanden (2005) (Period$_{kink} \approx \frac{2L}{C_k}$) as, $C_k$ = 21.6 km/s. Since the obtained kink speed is higher than the sound speed, the observed kink wave is in the fast mode. Since the CS density is obtained to be higher than its surroundings, Considering a density contrast of 5, in the low $\beta$ plasma regime one can estimate the Alfvén speed as $v_A$ = $c_k \sqrt{\frac{1+\rho_e/\rho_0}{2}}$ = 16.7 km/s, where $\rho_e$ and $\rho_0$ are the densities outside

12   *M. Ghiasi et al.*

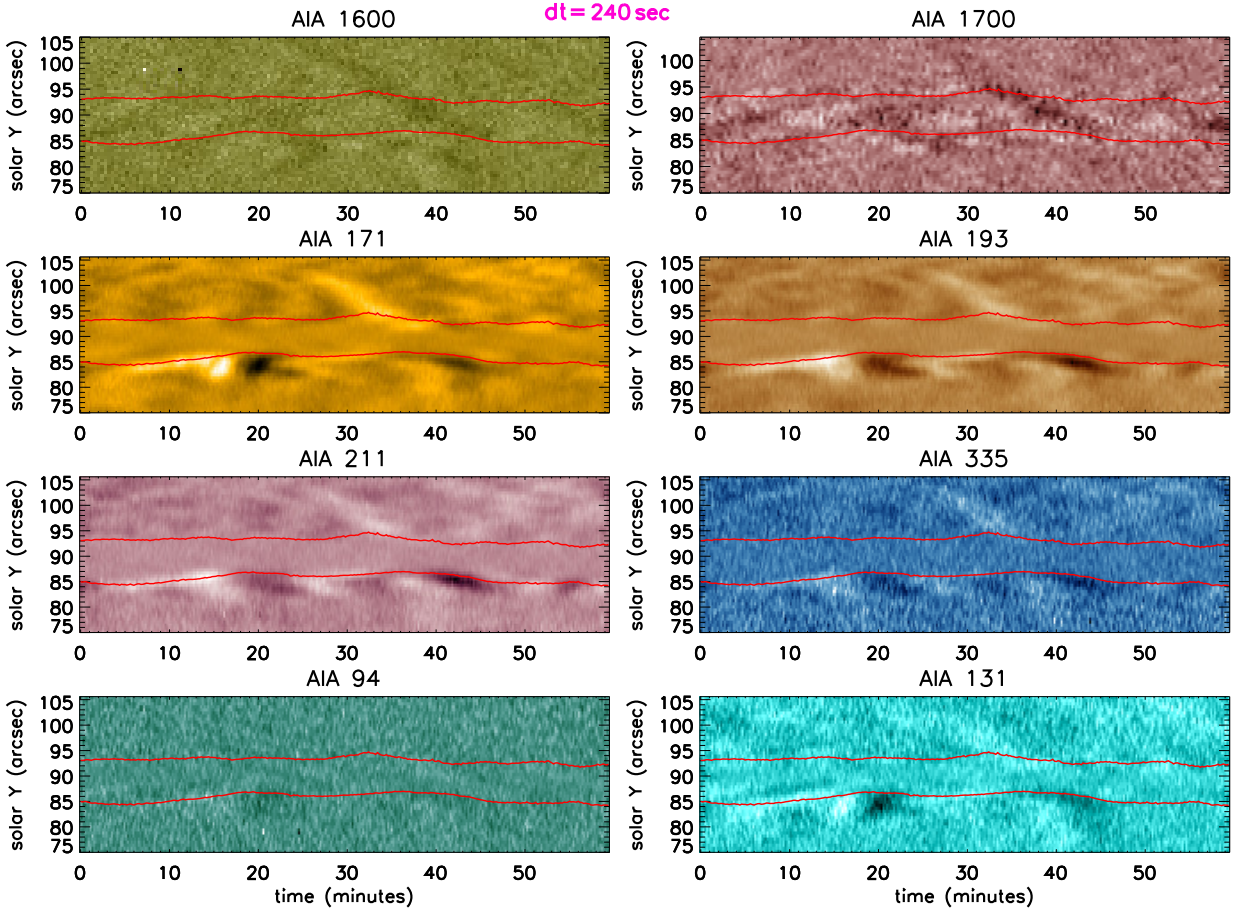

**Figure 13.** Running difference version of Fig. 12, with dt = 240 seconds. Red over-plotted lines on each panel are the intensity contours of AIA 193 Å channel, and selected in a way to outline the top and below borders of the CS structure.

and inside of the CS. Therefore, the magnetic field over the CS is obtained to be (B = $V_A \sqrt{4\pi n_e m_{Fe\ XII\ ion}}$ = 2.79[in Gauss unit]. The energy flux of the kink wave in the CS is estimated to be equal to ($F_A = \frac{1}{2} \rho v_{00}^2 V_A = n_e m_H v_{00}^2 V_A = 41.93$ W/m$^2$), where, $v_{00}$ is the transverse amplitude velocity of the kink wave.

• If we consider the observed magnetic field to be the final magnitude of the magnetic field at the ending of our study time ($B_{t_2}$ = 2.79 Gauss), then considering the Equ. 16 the initial magnetic field of the structure is estimated to be $B_{t_2}$ = 4.57 Gauss. Therefore, the rate of the decaying the magnetic field over the CS (as a result of its cross section expansion), $4.95 \times 10^{-4}$ Gauss per second.

**ACKNOWLEDGEMENTS**

The authors sincerely thank to the very constructive and useful comments of the unknown reviewer. They also thank the editors for providing a couple of revision time extensions. The authors thank Prof. Dr. Hossein Safari for his helpful suggestions. Hinode is a Japanese mission developed and launched by ISAS/JAXA, collaborating with NAOJ as a domestic partner, and NASA and STFC (UK) as international partners. Scientific operation of the Hinode mission is conducted by the Hinode science team organized at ISAS/JAXA. This team mainly consists of scientists from institutes in the partner countries. Support for the post-launch operation is provided by JAXA and NAOJ (Japan), STFC (U.K.), NASA, ESA, and NSC (Norway).

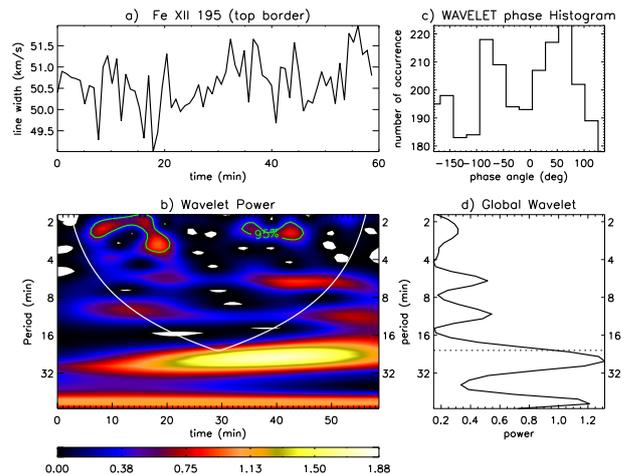

**Figure 14.** Panel a, shows the time series of the Fe XII 195.119 Å line widths (in km/s) over the CS top border. Panel b, and d represent the wavelet power spectrum, and global wavelet spectrum (summed over time), respectively. The light green contours outline the 95% confidence level. Area below the white solid line is the cone of influence (COI). Panel c demonstrates a histogram of the phase distribution.

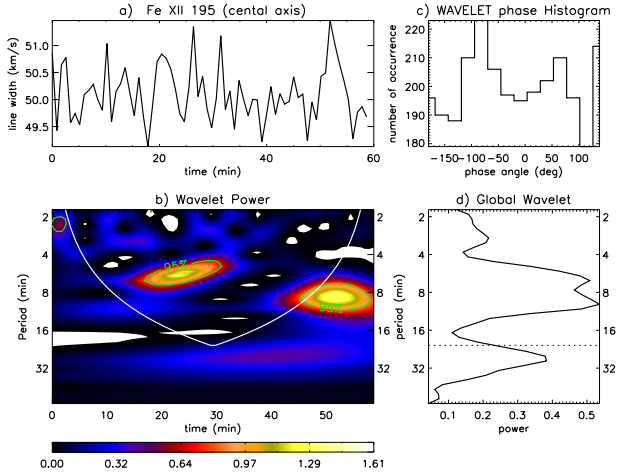

**Figure 15.** As 14, but for central axis of the CS.

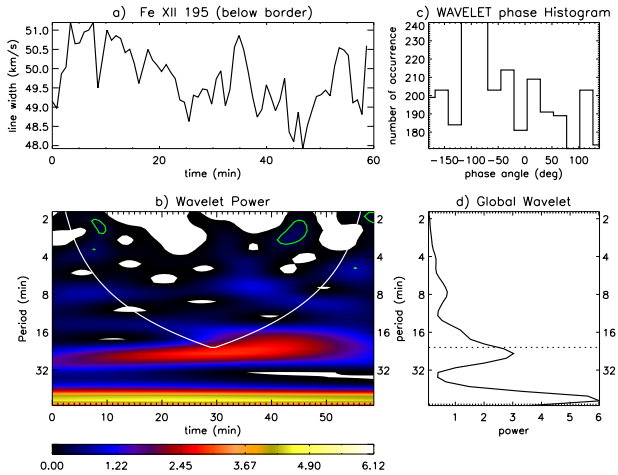

**Figure 16.** As 14, but for below border of the CS.

IRIS is a NASA small explorer mission developed and operated by LMSAL with mission operations executed at NASA Ames Research center and major contributions to downlink communications funded by ESA and the Norwegian Space Centre. Data are courtesy of NASA/SDO and the AIA and HMI science teams. Wavelet software was provided by C. Torrence and G. Compo, and is available at http://paos.colorado.edu/research/wavelets/.

## 5 DATA AVAILABILITY

The IRIS, EIS/Hinode, and AIA/SDO data underlying this article are publicly available in the following link (the NASA IRIS website): https://www.lmsal.com/hek/hcr?cmd=view-event&event-id=ivo://sot.lmsal.com/VOEvent%23VOEvent_IRIS_20140403_131610_3840259471_2014-04-03T13:16:102014-04-03T13:16:10.xml.

The data are taken during the campaign (IHOP 254. prominences.) with observation ID of OBS 3840259471. The wavelet software used for wavelet analysis is available at http://paos.colorado.edu/research/wavelets/.

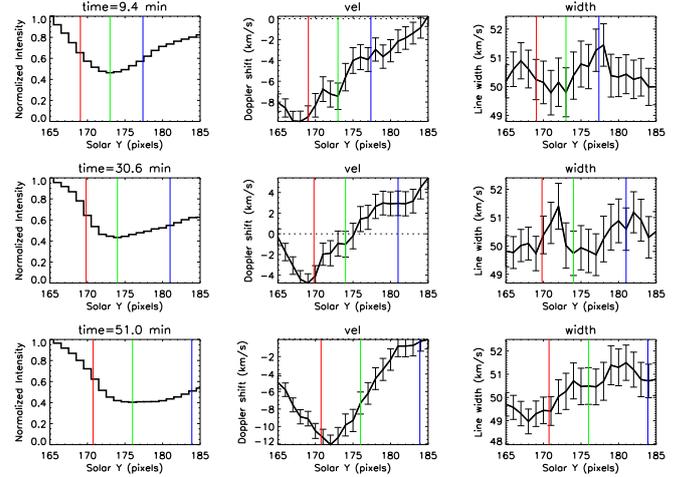

**Figure 17.** Comparison of Fe XII 195.119 Å intensity, Doppler shift, and line width at the moments where the CS thickness is maximum (pink arrows, Fig. 8). Red, green, and blue solid vertical lines denote the positions of the top border, central axis, and below border of the CS, at each studied frame time.

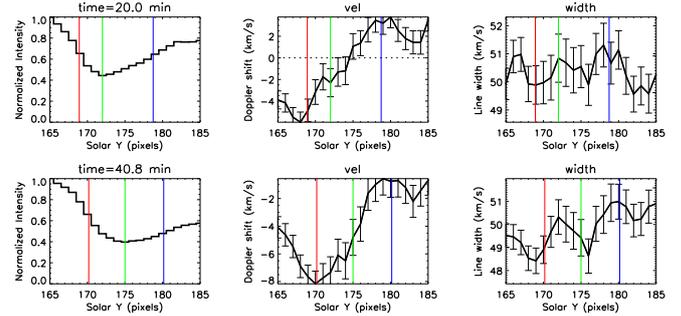

**Figure 18.** As Fig. 17, but for blue arrows demonstrated on Fig. 8.

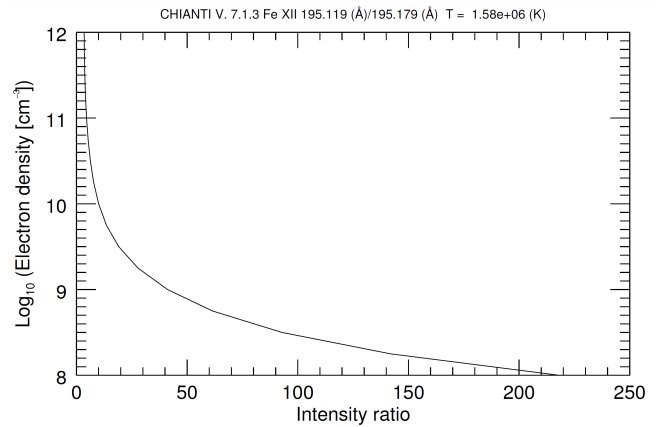

**Figure 19.** The electron density versus the line ratio of Fe XII 195.119(Å) / 195.179(Å). Version 7.1.3 of CHIANTI atomic database is used to plot this Figure. Courtesy of Levens et al. (2015).

14     *M. Ghiasi et al.*

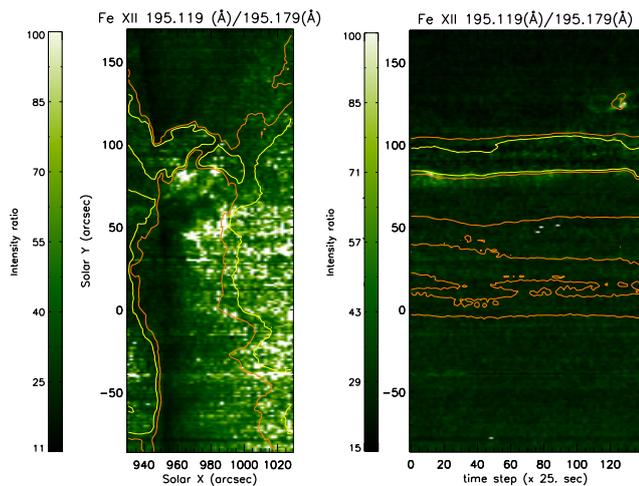

**Figure 20.** Left panel: Line ratio map of Fe XII 195.119(Å) / 195.179(Å). Right Panel: line ratio of Fe XII 195.119(Å) / 195.179(Å) in the EIS sit − and − stare mode. Yellow and orange contours demonstrate the intensities of 50000 and 60000 (erg/cm$^2$/s/str) for Fe XII 195.119(Å) line, respectively.

**REFERENCES**

Alzahrani F. M., Chen T., 1997, Real-Time Imaging, 3, 363–378
Antolin P., Shibata K., 2010, The Astrophysical Journal, 712, 494
Antolin P., Van Doorsselaere T., 2019, Frontiers in Astronomy and Space Science, 7
Antolin P., Schmit D., Pereira T. M. D., De Pontieu B., De Moortel I., 2018, The Astrophysical Journal, 856, 44
Aschwanden M., 2005, Physics of the Solar Corona. Praxis Publishing ublishing, Chichester, UK
Aschwanden M. J., Fletcher L., Schrijver C., Alexander D., 1999, The Astrophysical Journal, 520, 880–894
Banerjee D., Pérez-Suárez D., Doyle J. G., 2009, Astronomy and Astrophysics, 501, 15
Bendat J. S., Piersol A. G., 2000, Meas Sci Technol, 11, 1825–1826
Culhane J. L., Harra L. K., James A. M., et al. 2007, Solar Physics, 243, 19
De Moortel I., Nakariakov V. M., 2012, Royal Society of London, Philosophical Transactions Series A, 370, 3193
De pontieu B., McIntosh S. W., Carlsson M., Hansteen V. H., et al. 2007, Science, 318
Dere K., Landi E., Mason H., Monsignori Fossi B., Young P., 1997, Astronomy and Astrophysics Supplement Series, 125
Dorotovic I., Erdélyi R., Karlovsky V., 2008, IAU Symposium, 247
Erdélyi R., Fedun V., 2007, Science, 318, 1572
Gomez J. C. G., Jafarzadeh S., Wedmeyer S., Szydlarski M., Stangalini M., Fleck B., Keys P. H., 2020, Philosophical Transcations A, 379
Gonzalez R. C., Woods R. E., 1992, Digital Image Processing. Addison-Wesley Publishing Co. Inc.
Jess D. B., Mathioudakis M., Erdélyi R., et al. 2009, Science, 323, 1582
Jess D. B., Morton R. J., Verth G., et al. 2015, Space Science Reviews, 190, 161
Jess D. B., Jafarzadeh S., Keys P. H., Stangalini M., G. V., Grant S. D. T., 2023, Living Reviews in Solar Physics, 20, 1:170
Korendyke C. M., Brown C. M., Thomas R. J., et al. 2006, Applied Optics IP, 45, 8674
Kukhianidze V., Zaqarashvili T. V., Khutsishvili E., 2006, Astronomy and Astrophysics, 449
Laigong G., Sitong W., 2023, Applied Sciences, 13
Landi E., Del Zanna G., Young P., Dere K., Mason H., 2012, Astronomy and Astrophysics Supplement Series, 744
Lemen J. R., Title A. M., Akin D. J., et al. 2012, Solar Physics, 275, 17
Levens P. J., Labrosse N., Fletcher L., B. S., 2015, Astronomy and Astrophysics, 582
Lynn N. D., A Sourav I., Santoso A. J., 2021, IOP Conference Series: Materials Science and Engineering, 1096, 012079
Mathioudakis M., Jess D. B., Erdeĺyi R., 2012, Space Science Reviews, pp 1–17
McIntosh S. W., De Pontieu B., Carlsson M., et al. 2011, Nature, 475, 477
Morton R. J., Verth G., Hillier A., Erdélyi R., , The Astrophysical Journal, 748
Morton R. J., Verth G., Fedun V., Shelyag S., Erdélyi R., , The Astrophysical Journal, 768
Morton R. J., Erdélyi R., Jess D. B., Mathioudakis M., 2010, The Astrophysical Journal Letters
Morton R. J., Verth G., Jess D. B., Kuridze D., Ruderman M. S., Mathioudakis M., Erdélyi R., 2012, Nature Communications, 1315
Morton R. J., Tomczyk S., Pinto R., 2015, Nature Communications, 6, 7813
Nakariakov V. M., Verwichte E., 2005, Living Reviews in Solar Physics, 2, 65 pp.
Nakariakov V. M., Ofman L., Deluca E., Roberts B.and Davila J., 1999, Science, 285, 862–864
Nakariakov V. M., Anfinogentov S. A., Antolin P., Jain R., Y. K. D., 2021, Space Science Reviews, 217
Ramnarayan Saklani N., Vasundhara V., 2019, International Journal of Computer Applications, 178
Ruderman M. S., Shukhobodskaya D., Shukhobodskiy A. A., 2019, Frontiers in Astronomy and Space Science, 6
Su W., Guo Y., R. E., Ning Z. J., Ding M. D., Cheng M. D., Tan B. L., 2018, Scientific Reports, 8
Terradas J., Goossens M. Verth G., 2010b, Astronomy and Astrophysics, 524
Torrence C., Compo G. P., 1998, Bull. Am. Meteorological Soc., 79, 61
Van Doorsselaere T., Nakariakov V., E. V., 2008, The Astrophysical Journal, 676
Verth G., Jess D. B., 2015, Reviews in Geophysics
Zaqarashvili T. V., 2003, Astronomy and Astrophysics, 399
Zaqarashvili T. V., Khutsishvili E., Kukhianidze V., Ramishvili 2007, Astronomy and Astrophysics, 474, 627

This paper has been typeset from a T<sub>E</sub>X/L<sup>A</sup>T<sub>E</sub>X file prepared by the author.